\documentclass[aip,amsfonts,amsmath,amssymb,reprint]{revtex4-1}
\usepackage{graphicx}
\usepackage{dcolumn}
\usepackage{bm}

\usepackage{xcolor}
\usepackage[english]{babel}
\usepackage[utf8]{inputenc}
\usepackage[T1]{fontenc}
\usepackage{mathptmx}
\usepackage{tabularx,booktabs}
\usepackage{soul}
\usepackage{color}
\usepackage{soul, xcolor}
\setstcolor{red}
\usepackage{kotex}

\usepackage{lipsum}
\usepackage{hyperref}

\usepackage{amsmath}

\begin{document}

\title{Machine learning approaches for Kuramoto coupled oscillator systems}
\author{Je Ung Song}
\author{K. Choi} 
\affiliation{CCSS and CTP, Department of Physics and Astronomy, Seoul National University, Seoul 08826, Korea}
\author{B. Kahng}
\email{bkahng@kentech.ac.kr}
\affiliation{Center for complex systems, KI for Grid Modernization, Korea Institute of Energy Technology, Naju, Jeonnam 58217, Korea}

\date{\today}

\begin{abstract}
Recently, there has been significant advancement in the machine learning (ML) approach and its application to diverse systems ranging from complex to quantum systems. As one of such systems, a coupled-oscillators system exhibits intriguing collective behaviors, synchronization phase transitions, chaotic behaviors and so on. Even though traditional approaches such as analytical and numerical methods enable to understand diverse properties of such systems, some properties still remain unclear. Here, we applied the ML approach to such systems particularly described by the Kuramoto model, with the aim of resolving the following intriguing problems, namely determination of the transition point and criticality of a hybrid synchronization transition; understanding network structures from chaotic patterns; and comparison of ML algorithms for the prediction of future chaotic behaviors. The proposed method is expected to be useful for further problems such as understanding a neural network structure from electroencephalogram signals.      
\end{abstract}

\maketitle

\begin{quotation}
The machine learning approach is useful in understanding diverse phenomena in coupled oscillator systems, particularly those that cannot be inferred using traditional analytical and numerical methods. Herein, we consider two problems of the Kuramoto model, namely a hybrid synchronization transition that occurs in fully connected networks and understanding the network structure from a time-evolution chaotic pattern generated by Kuramoto oscillators in an unknown structured network. In the case of the latter, the success rate of the matching of elements in the adjacency matrices between a testbed network and the inferred network is estimated as $96.4\pm 0.9\%$.   
\end{quotation}

\maketitle

\section{Introduction}

Recently, the incorporation of the machine learning (ML) approach has facilitated progressive advances in diverse fields of science and engineering. Herein, we focus on dynamical systems that exhibit synchronization transitions and chaotic patterns. Chaotic behaviors are observed in a variety of systems in nature, such as the cardiac cycle, neuroscience, climate, and the stock market. It is interesting to explore chaotic signals in such systems for predicting their dynamical evolution. In a traditional approach, this is achieved by setting up an appropriate model for the current chaotic pattern, which is then simulated to predict the dynamical evolution. However, the ML approach using reservoir computing~\cite{maass2002real,Jaeger2004,Lukosevicius2009} identifies the underlying factors of the current chaotic behavior and then provides a model-free prediction of the dynamical evolution~\cite{Lu2017,Carroll2018,Lu2018,Pathak2017,Pathak2018,Weng2019,Jiang2019,Fan2020,Lai2020}. 

Chaotic patterns are generated not only by single-particle nonlinear dynamic equations, but also through the cooperation of multiple elements in a system. It may be interesting to understand how these elements are interwoven and cooperate in a system \cite{Nitzan2017,wang2016,eroglu2020}. For instance, in neurophysiology, research on classifying and capturing physiological events, such as seizures, strokes, or headaches, has been conducted by identifying the correlations among electroencephalogram (EEG) signals. Recently, ML approaches have accelerated the progress of identifying the cooperation among neuronal components~\cite{mormann2005predictability, mirowski2009classification, chandaka2009cross, williamson2012seizure}. 

Synchronization of chaotic patterns induced by the cooperation of multiple elements in such neural systems may signal a stable state. Hence, a phase transition from a disordered state to an ordered state naturally arises as an interesting issue in such complex systems. Given the recent advances in ML algorithms for the studies of phase transitions in equilibrium and nonequilibrium systems such as the Ising model and percolation~\cite{Bohrdt2019,Zhang2019,Carrasquilla2017,Venderley2018,Beach2018,Ni2019,Broecker2017}, it would be interesting to consider the synchronization transition using the ML approach. 

As a model for describing the behaviors of coupled oscillators, the Kuramoto model (KM)~\cite{Kuramoto1975,Kuramoto1984} may be a suitable candidate to deal with the above-mentioned issues simultaneously. This system exhibits not only the chaotic dynamics provoked by nonlinear couplings between oscillators but also different types of synchronization transitions depending on the underlying connection topologies or variations in the model~\cite{martens2009exact,pazo2009existence,skardal2018low,Tang2011,Rodrigues2016,Pazo2005,Basnarkov2007,Coutinho2013,Song2020,Choi2013,Yoon2015,Rodrigues2016}. 
In addition to the conventional transition types, namely first- and second-order transitions, a hybrid synchronization transition (HST) can be considered as an appropriate model choice wherein the properties of both the first- and second-order transitions can be observed at the same transition point ~\cite{Pazo2005,Basnarkov2007,Coutinho2013,Song2020}. In this case, the determination of the transition point is challenging owing to the large fluctuations over different configurations. Recently, for the hybrid percolation transition, a transition point was determined by a unusual method~\cite{rer}. Moreover, scaling behavior has not been explored because of the challenges faced in large-scale numerical simulations. 

In this paper, we first consider a second-order and a HST of the KM: the determinations of a transition point and the correlation length exponent $\bar{\nu}$ for each case. Note that the correlation length exponent of the HST for the model we consider here has not been known yet. We demonstrate that the exponent value can be determined easily as much as we did for the second-order synchronization transition. Thus, the ML approach, specifically the fully-connected neural network (FCN) provides transition-type-free facilitation. Next, we consider the inference of network structure from chaotic phases of each oscillator of the KM by applying a deep learning algorithm.

The remainder of this paper is organized as follows: In Sec.~\ref{sec:MLapproaches}, we briefly introduce the ML algorithms we use in the paper. In Sec.~\ref{sec:SyncTransition}, a snapshot of the phases for all oscillators is considered to discriminate between synchronized and asynchronized states and determine the transition point of the KM with a finite size. In addition, we verify the scaling exponent $\bar{\nu}$ of the KM exhibiting a second-order transition and determine $\bar{\nu}$ for a KM with a degree-frequency correlation exhibiting a hybrid transition~\cite{Coutinho2013}. In Sec.~\ref{sec:reconstruction}, given the credible results obtained by learning phase dynamics, the underlying structure of the mouse visual cortex network is reconstructed based on the phase dynamics of all oscillators using a well-trained machine. Finally, the conclusions are presented in Sec.~\ref{sec:conclusion}. In Appendix, we consider prediction horizon of an evolving chaotic pattern produced by the  KM. In this study, we adopt not only the reservoir computing (RC) algorithm conventionally applied, but also other algorithms such as classical recurrent neural network (RNN), convolution neural network (CNN), and FCN. Therefore, we could compare their efficiencies for the prediction of future chaotic patterns of the KM.

\section{Machine learning approaches}\label{sec:MLapproaches}
\subsection{Artificial neural networks}\label{subsec:anns}

A FCN comprising fully connected layers (FLs) represents the basic structure of a feed-forward neural network (FNN). Neurons in each FL are connected to all the input components in the preceding layer, and all such connections between two consecutive layers are represented by the weight matrix $\mathbf{W}$. For a given input of $\mathbf{x}$, the states of neurons $\mathbf{y}$ are updated by the following equation: 
\begin{align}
	\mathbf{y} = f(\mathbf{W}\mathbf{x}+\mathbf{b}),
	\label{eq:ann_eq}
\end{align}
where $\mathbf{b}$ denotes the bias exerted on the neurons, and $f(\mathbf{x})$ is the activation function. In general, nonlinear functions such as sigmoid, tanh, ReLU, and softmax are used as the activation functions, and one can select different forms of the activation function for each layer. In the FNN, neuron states serve as the input for the subsequent layer until they reach the final output layer.

By combining FLs, convolutional layers (CLs), and pooling layers, a convolutional neural network (CNN) can be constructed. By passing through the CL, the input data are transformed by filters in the CL, which are useful for maintaining spatial information and identifying the spatial patterns of the input, such as translational symmetry and rotational symmetry.

As another deep learning model for dealing with sequential data, recurrent neural networks (RNNs) comprise recurrent layers with a cyclic connection topology, which distinguishes them from FCNs. This RNN structure resembles biological brain modules, which also exhibit recurrent connection pathways. 

Reservoir computing (RC), a class of recurrent neural networks, is composed of an input layer, an output layer, and a reservoir layer that connects the input and output layers. Neurons in the reservoir layer comprise internal links, including a self-loop, as in the recurrent layer.

In this study, as the input vector $\mathbf{u}(t)$ goes in, the state vector $\mathbf{r}$ is updated according to the equation
\begin{align}
	\mathbf{r}(t+1) = (1-\lambda)\mathbf{r}(t) + \lambda \tanh \left( \mathbf{Ar}(t)+ \mathbf{W}_{\textrm{in}} \begin{bmatrix} b_{\textrm{in}} \\ \mathbf{u}(t) \end{bmatrix} \right),
	\label{eq:RC_state}
\end{align}
where $\mathbf{A}$ denotes the weighted adjacency matrix of the reservoir network, $\mathbf{W}_{\textrm{in}}$ denotes a random matrix that maps an input vector $\mathbf{u}(t)$ to a state vector $\mathbf{r}(t)$, $\lambda$ denotes the leakage rate, and $b_{in}$ represents the bias term. The output vector $\mathbf{y}(t)$ is determined by a linear function: 
\begin{align}
	\mathbf{y}(t) = \mathbf{W}_{\textrm{out}} \begin{bmatrix} b_{\textrm{out}} \\ \mathbf{u}(t) \\ \mathbf{r}(t) \end{bmatrix}
	\label{eq:RC_output}
\end{align}
where $\mathbf{W}_{\textrm{out}}$ denotes the output matrix that maps a reservoir state to an output vector for a given bias of $b_{out}$.

\subsection{Supervised learning}\label{subsec:supervised}

Although a variety of structures can be designed, FNN $\mathcal{F}$ is determined by model parameters $\{w \}$, including weight and bias. The model produces output $\mathbf{y}$ for a given input $\mathbf{x}$, which is expressed in the functional form: 
\begin{align}
	\mathbf{y} = \mathcal{F}\left[\{w\}\right](\mathbf{x}) \,.
	\label{eq:supervised}
\end{align}
In supervised learning, the model parameter $\{w\}$ of the FNN is adjusted such that the output $\mathbf{y}$ is close to the desired output $\bar{\mathbf{y}}$ according to the given input $\mathbf{x}$ of the training dataset. To minimize the difference between $\mathbf{y}$ and $\bar{\mathbf{y}}$, or the cost (loss, energy) function $E$, determined by the root mean square, mean absolute, or cross entropy, the model parameter $w$ is tuned using the gradient descent method, which is the fundamental method for training FNNs including RNNs:
\begin{align}
	w_{\textrm{new}} = w - \alpha \partial_{w}E(\mathbf{y},\bar{\mathbf{y}})
	\label{eq:gdm}
\end{align}
where $\alpha$ denotes the learning rate. With random initial values of $\{w\}$, a well-trained FNN is obtained by repeating the learning process of Eq.~\eqref{eq:gdm}~\cite{Goodfellow2016}.

In the case of RC, when the updated state vector $\mathbf{r}$ of the reservoir is given by Eq.~\eqref{eq:RC_state}, the output weights $\mathbf{W}_{\textrm{out}}$ are determined using equation~\cite{Lukosevicius2009,Weng2019}
\begin{align}
	\mathbf{W}_{\textrm{out}} = \mathbf{Y}\mathbf{X}^{\top}(\mathbf{X}\mathbf{X}^{\top}+\gamma \mathbb{I})^{-1}
	\label{eq:RC_weight}
\end{align}
where $\gamma$ denotes the ridge regularization parameter and $\mathbb{I}$ denotes an identity matrix. Additionally, $\mathbf{X}$ and $\mathbf{Y}$ represent the collecting matrices of the state vector $[ b_{out}, \mathbf{u}(t), \mathbf{r}(t) ]^{\top}$ and the desired output vector $\bar{\mathbf{y}}(t)$ in the training process, respectively. While the classical RNN adopts back propagation through time, which is based on a gradient descent method for recurrent layers, for RC, $\mathbf{A}$ and $\mathbf{W}_{\textrm{in}}$ are randomly created and unchanged during training, and only the output weights $\mathbf{W}_{\textrm{out}}$ are computed.

\section{Synchronization transitions}\label{sec:SyncTransition}

Synchronization is a macroscopic-scale collective pattern generated from each oscillator of the KM. The Kuramoto model comprises $N$ globally coupled oscillators interacting with each other via nonlinear coupling, and can be defined as
\begin{align}
	\dot{\theta}_i = \omega_i+\frac{K}{N}\sum_{j=1}^N \sin (\theta_j-\theta_i),
	\label{eq:KM}
\end{align}
where the dot on $\theta_i$ indicates the derivative of phase $\theta_i$ of oscillator $i$ with respect to time; $\omega_i$ denotes the natural frequency of oscillator $i$, which follows the distribution $g(\omega)$; and $K$ denotes the coupling strength. The collective behavior of the system is quantified by the complex order parameter $Z$, which is defined in the limit $t \to \infty$ as 
\begin{align}
	Z = re^{i\psi}=\frac{1}{N}\sum_{j=1}^N e^{i\theta_j},
	\label{eq:order_parameter}
\end{align}
where $r$ denotes the order parameter measuring the extent of phase coherence, $\psi$ denotes the average phase angle. When $K$ is small, $r$ is zero in the limit $N\to \infty$.  As $K$ is increased, $r$ approaches a nonzero value at a transition point $K_c$ in the limit $N \rightarrow \infty$, which implies the occurrence of global phase synchronization. For finite systems of size $N$, the transition point depends on $N$, denoted as $K_c(N)$. Hereafter, we use the normalized coupling strength $J \equiv K/K_c$. Hence, the transition point is $J_c = 1$.

Depending on the shape of $g(\omega)$, the synchronization transition has three types: i) When $g(\omega)$ is uni-modal, the transition is of second-order, so that the order parameter increases continuously as $r\sim (J-J_c)^\beta$ for $J \ge J_c$. ii) When $g(\omega)$ is bi-modal, the transition is of first-order, so that the order parameter jumps to a finite value at $J=J_c$. iii) When $g(\omega)$ is finite in the interval $[-\omega_0, \omega_0]$ and zero in other region, the transition is hybrid and the order parameter is expressed as $r-r_0 \sim (J-J_c)^\beta$. iv) A HST also occurs for a particular case that oscillators locate on scale-free networks with degree exponent $\gamma=3$. A scale-free network is a network with heterogeneous numbers of connected oscillators $\{k_i\}$ of each oscillator $i$. They have a power-law distribution $P_d(k)\sim k^{-\gamma}$. A HST occurs when $\omega\sim k_i$. For this case, even though analytical solution of the HST is present, the correlation length exponent $\bar{\nu}$ is still unknown because of the difficulty of numerical simulations. Here, we determine the exponent $\bar{\nu}$ as a target of the ML approach.    

Herein, using a fully connected neural network (FCN), we train the snapshots of the phases of each oscillator in subcritical and supercritical regimes, respectively, and eventually identify the transition point of the system with a finite size and its changes with varying system size. Here, we consider two different systems i) and iv) for the utilization of NN approaches for synchronization transition. The case i) is to check whether the ML methodology is correct and the case iv) is to determine the unknown critical exponent. 

\subsection{Second-order synchronization transition}\label{subsec:secondorder}

Here, $g(\omega)$ is considered as a normal distribution,
\begin{align}
g(\omega) = \frac{1}{\sqrt{2\pi}} e^{-\frac{\omega^2}{2}} \,.
\label{eq:frequency}
\end{align}
In this distribution, a second-order synchronization transition occurs at $K_c = 2/[\pi g(0)] = \sqrt{8/\pi}$ in the limit $N \rightarrow \infty$~\cite{Kuramoto1975,Kuramoto1984}. The order parameter for $J \ge J_c$ is expressed as $r\sim (J-J_c)^\beta$.

\begin{figure*}[!t]
\centering
\includegraphics[width=0.98\linewidth]{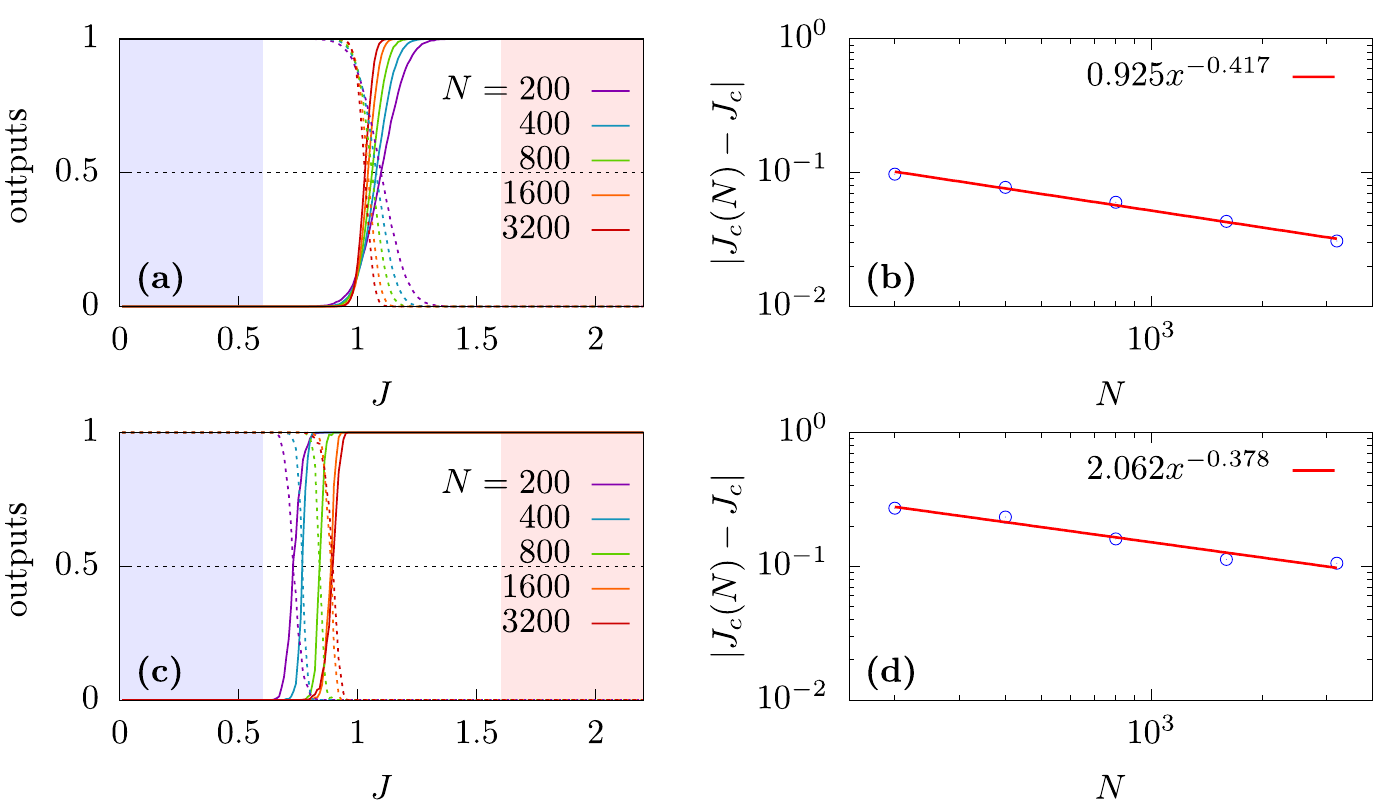}
\caption{(a) Outputs are the probabilities for the system to be in the state (0,1) and (1,0) in the subcritical and supercritical regions, respectively. Plot of outputs versus the coupling strength $J$ for the Kuramoto model with a system size of $N=200, 400, 800, 1600$, and $3200$ and natural frequencies following the normal distribution $g(\omega)$. The intersection point of the two output lines indicates $J_c(N)$ for a given $N$. (b) Behavior of $|J_c(N) -J_c|$ with increasing $N$. The straight line represents the fitting line with a slope of $-0.417 \pm 0.021$. (c) Similar plot to that of (a) for the Kuramoto model with the degree-frequency correlation on a scale-free network with $\gamma=3$. (d) Similar plot to that of (b) for the Kuramoto model considered in (c). The straight line represents the fitting line with a slope of $-0.378 \pm 0.021$.
\label{fig:fig1}
}
\end{figure*}

This system is trained to distinguish between a synchronous state and an asynchronous state using the phases of all oscillators as the input for the NN. The fourth-order Runge-Kutta method with a time step of $\delta t=0.01$ is adopted to generate $2 \times 10^4$ datasets of $\{ \theta _i \}$ for each given value of $N$ and $J \in [0.01, 2.20]$ with $\delta J=0.01$. For each configuration, the sets of natural frequencies, $\{ \omega_i \}$, are randomly selected from the normal distribution given in Eq.~\eqref{eq:frequency} and initial phases, $\{ \theta_i(t=0) \}$ are selected randomly from the range of $[0, 2\pi]$. To avoid any transient behavior, we collect the snapshots of the phases, $\{ \theta _i \}$, after the first $10^6$ steps. As the inputs, we use data preprocessing by taking the cosine and sine for each phase  owing to the cyclic feature of $\theta_i$, which contains a $2\pi$ periodicity. 

For training datasets, each snapshot is labeled through one-hot encoding, where the configurations obtained in the subcritical region of $J \in [0.01, 0.6]$ are encoded as $(0, 1)$, and those in the supercritical region of $J \in [1.6, 2.2]$ are encoded as $(1, 0)$. Note that the test region $[0.6, 1.6]$ is asymmetric with respect to the transition point $J_c=1$.  

Constructing an FCN and CNN, we train the NNs with labeled snapshots of $N$ phases $\{\cos \theta _i, \sin \theta_i \}$. When the network is optimized after training, the snapshots generated in the entire region of $J$ are input at the test stage. The trained NN produces two outputs $(0,1)$ and $(1,0)$ representing the probabilities for the system to be in the subcritical and supercritical regions, respectively, as depicted in Fig.~\ref{fig:fig1}(a). The intersection point of the two output curves indicates a transition point $J_c(N)$ for a given system size $N$. Because $J_c(N)$ approaches the critical point $J_c=1$ as $N$ increases, we can determine the value of the critical exponent $\bar{\nu}$ using the relation $|J_c(N)-J_c| \sim N^{-1/\bar{\nu}}$. Fig.~\ref{fig:fig1}(b) depicts a finite size scaling of $J_c(N)$ with the exponent $1/\bar{\nu}=0.417 \pm 0.021$ and thus $\bar{\nu}=2.404\pm 0.121$, which is reasonably in agreement with the analytically solved value $\bar{\nu} = \bar{\nu}' = 5/2$ for the KM with natural frequencies of each oscillator following the Gaussian distribution. Thus, we verify that the ML approach is successful to determine the correlation length exponent for the synchronization transition of the KM. 

\subsection{Hybrid synchronization transition}\label{subsec:hybrid}
After verifying the applicability of the ML approach for understanding the critical behavior of a second-order synchronization transition, here, we apply it to the model iv) with a HST. As $J$ is increased to $J_c=1$ from the subcritical regime, the order parameter jumps at $J_c$, and then increases gradually beyond $J_c$. Therefore, the order parameter is discontinuous and also has critical properties. The KM iv) with the degree-frequency correlation on scale-free networks exhibits a second-order (first-order) synchronization transition when the exponent of degree distribution $\gamma > 3$ ($2<\gamma<3$) under a unimodal $g(\omega)$ distribution. When $\gamma=3$, however, the KM exhibits a HST~\cite{Coutinho2013}. In general, for such HSTs, it is non-trivial to determine transition points of finite systems because of strong sample-to-sample fluctuations of transition points. Here, we demonstrate that using the ML approach, we determine transition points of HSTs for finite systems with different system sizes, and obtain the correlation length exponent $\bar{\nu}$.   

By training the NN with phase snapshots of oscillators on this system following the same way as in the second-order transition case, we obtain two output lines to evaluate $\bar{\nu}$ for this HST case. As depicted in Figs. 1(c) and 1(d), $|J_c(N)-J_c|$ scales as $N^{-0.378}$. This implies that $1/\bar{\nu} \approx 0.378$.

\section{Inference of network structure}\label{sec:reconstruction}

A network structure may be inferred through time-evolution data in diverse fields. If the time-evolution patterns of two nodes are positively (negatively) correlated, then the two nodes are regarded as being connected by an excitatory (inhibitory) link. For example, a modular structure could be identified by observing the synchronized pattern created by the Kuramoto model among a group of oscillators~\cite{Oh,Arenas2006}. Moreover, a hierarchical structure of modules could be recognized. Recently, local time series information enables to infer global temporal structure~\cite{Kim_2021}. Here, we consider the reconstruction of a network structure using the ML algorithm by learning the data of nonlinear oscillators of the KM. Identifying the network topology is one of the main problems in predicting the behavior of the system and understanding the interactions among individuals or implicit mechanisms in various systems, such as neuronal connections in the brain and epidemics in social networks. Because it is difficult to directly identify neuronal networks, indirectly recovering a network through the time-evolution data of nodes has been attempted. Applying this approach to networks that are more general than the modular network, we assume a situation wherein the connections of the network are not provided but only the individual patterns produced through inherent interactions between them are available. As it has significant application potential, the ML approach is adopted to detect the entire network topology by comprehending the interactions among individual patterns.

For this purpose, we generated $10^6$ training datasets for the coupled oscillators governed by the KM on Erd\H{o}s-R\'enyi (ER)-type random networks, 
\begin{align}
\dot{\theta}_i(t) = \omega_i+K\sum_{j=1}^N A_{ij} \sin (\theta_j-\theta_i),
\label{eq:KM_net}
\end{align}
where $A_{ij}$ represents the adjacency matrix of a given network. We consider $N=29$ because the size of the target network~\cite{nr} is 29. The natural frequency set $\{ \omega_i \}$ is selected regularly from the Gaussian distribution, as given in Eq.~\eqref{eq:frequency}. The set $\{ \omega_i \}$ is assigned randomly. The initial phases $\{\theta_i(0)\}$ are assigned randomly in the range of $[-\pi, \pi]$ for all $i$. Using the fourth-order Runge-Kutta method with a time step $\delta t = 0.05$ up to a total of $200$ steps, sets of time series of phases $\{ \theta_i (t)\}$ are generated, which are used as the input for training an NN. Considering a generated set of $\{ \theta_i(t)\}$ as the input and the given network as the target output, the NN is optimized and regarded as a trained NN. 

\begin{figure}[!t]
\centering
\includegraphics[width=0.99\linewidth]{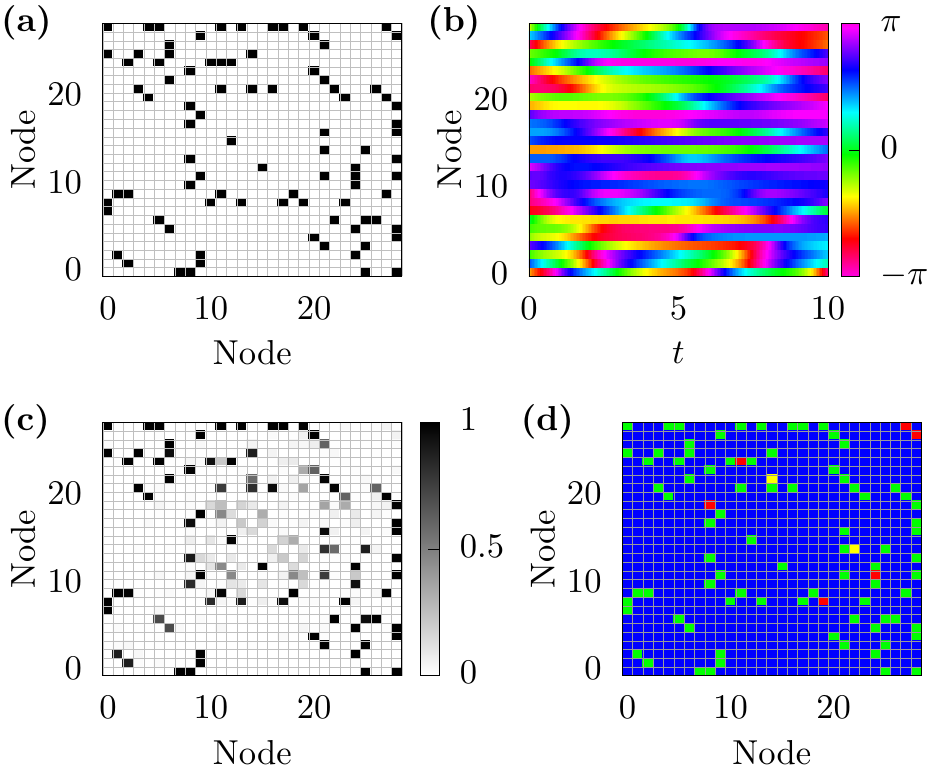}
\caption{Reconstruction of the visual cortex network with $N=29$. (a) Actual adjacency matrix of the visual cortex network. The black squares indicate the links existing on the network. (b) Input dataset of the actual phase evolution, $\{ \theta_i (t) \}$, for one of the sample sets of random natural frequencies. (c) Outputs obtained using the RNN. The output elements are real-valued in the range of $[0, 1]$. (d) Comparison between the actual adjacency matrix and rounded values of the obtained output elements in (c). The blue (green) square represents the element in the adjacency matrix with a value of zero (one) in the actual network and that of zero (one) in the rounded output. The red (yellow) square represents the element in the adjacency matrix with a value of zero (one) in the actual network and that of one (zero) in the rounded output.
\label{fig:fig2}
}
\end{figure}

Next, we examine the performance of the trained NN using the test datasets obtained. On the mouse visual cortex network~\cite{nr}, whose adjacency matrix is depicted in  Fig.~\ref{fig:fig2}(a), we assign natural frequencies $\{\omega_i \}$ similar to that in the previous method, and run the Kuramoto dynamics given by Eq.~\eqref{eq:KM_net}. With different sets of natural frequencies, $10^3$ patterns of test datasets $\{\theta_i (t)\}$ are generated. A sample is illustrated in Fig.~\ref{fig:fig2}(b). These datasets are used as the input sets of the trained NN. For each input dataset, the product of the trained NN contains $29\times 29$ real numbers in the range of $[0, 1]$ (see Fig.~\ref{fig:fig2}(c)). Each of them  represents the occupation probability of a link in the adjacency matrix, and are rounded off to 0 or 1. These values construct the adjacency matrix of an inferred network, for instance, the adjacency matrix depicted in Fig.~\ref{fig:fig2}(d), where the element with a value of $0$ in the real network and that of $0$ in the inferred network is denoted in blue; that with a value of $1$ and $1$ in the real and inferred networks is indicated in green; that with a value of $1$ and $0$ is depicted in red, and that with a value of 0 and 1 is depicted in yellow. The fraction of elements in blue and green represents the success rate of the inferred network in our testbed, which is estimated as $96.4\pm 0.9\%$ over all trials with the $10^3$ input datasets. 

We generated $10^6$ ER networks for training the neural network model and applied it to the cortex network for evaluation of the performance. The cortex network structure is successfully reproduced. Based on this performance, one can think that reconstructions of other ER networks would not be different.
We note that in this paper, we proposed a deep learning algorithm for the inference of a network structure from a chaotic pattern. This algorithm differs from other ML algorithms used in previous studies of network inference~\cite{Nitzan2017,eroglu2020}. So it is interesting to compare our algorithm with others in diverse perspectives. In-depth studies on the network inference remain as future works.

\section{Conclusion}\label{sec:conclusion}

To summarize, we employed the ML approach for coupled oscillator systems to classify the types of synchronization transitions and to perform model-free prediction for future phase dynamics by exploiting the chaotic properties of the system. We demonstrated that the scaling behavior of the system is not only verified for the second-order synchronization transition between an asynchronous state and a synchronous state, but also for the KM exhibiting a HST in which numerical analysis for finite-size scaling is challenging. Furthermore, despite the nonlinearity of the system, we successfully predicted the future behavior of the phase dynamics by employing ML approaches. A classical RNN method seems to be more efficient than the RC method. Verification of the learning of the chaotic dynamics of coupled oscillators encourages the training of ANNs with the patterns of individuals on a real brain network. Underlying connections between the patterns can be identified using a well-trained machine, and this approach can be extended to other problems for detecting the topology of a system. Additionally, such a model-free prediction of nonlinear dynamics suggests that such ML methods can overcome the disadvantages of analysis with modeling and simulations and can be extensively applied to other nonlinear models or systems in nature. Consequently, we expect that our study will accelerate the employment of ML in nonlinear/chaotic systems with multiple elements.

\begin{acknowledgments}
This work was supported by the National Research Foundation of Korea by Grant No. NRF-2014R1A3A2069005 and KENTECH Research Grant (KRG2021-01-007). \\
\end{acknowledgments} 

\noindent {\bf Data availability}: The code is available to the public at https://github.com/ckj0721/MLKM.

\appendix

\section{Comparison of the efficiencies to forecast phase evolutions by several ML algorithms}\label{sec:PhaseEvolution}


Recently, time-series data have been produced abundantly from social and natural systems and are easily accessible. Some examples of time-series data are the daily closing stock prices, daily highest temperatures, and heights of ocean tides. These time-series data are analyzed by extracting meaningful statistics using the auto-correlation function between two different times, enabling the prediction of future dynamical behavior.  Such data are produced by many elements through nonlinear and complex interactions, whose patterns may be chaotic. One such example is the dataset of the evolving phases of Kuramoto oscillators in an asynchronized state. In such cases, the extraction of a hidden pattern from the time-series data can rarely be achieved using traditional approaches. However, recently, a RC method of ML approaches has been applied to the ergodic time-series data of the R\"ossler system, Lorenz system, and spatiotemporally chaotic Kuramoto-Sivanshinsky equation~\cite{Lu2017}. ML can replicate chaotic attractors and calculate Lyapunov exponents from data~\cite{Pathak2017}. The chaotic pattern obtained using the RC method is similar to that directly obtained from each system. Here, we apply diverse ML algorithms to the Kuramoto model in an asynchronized state and obtain the future time-evolution patterns of the phases, and aim to compare the accuracies of the ML algorithms. We use FCN, CNN, and RNN methods to predict the Kuramoto system.

\begin{figure}[!h]
\centering
\includegraphics[width=0.98\linewidth]{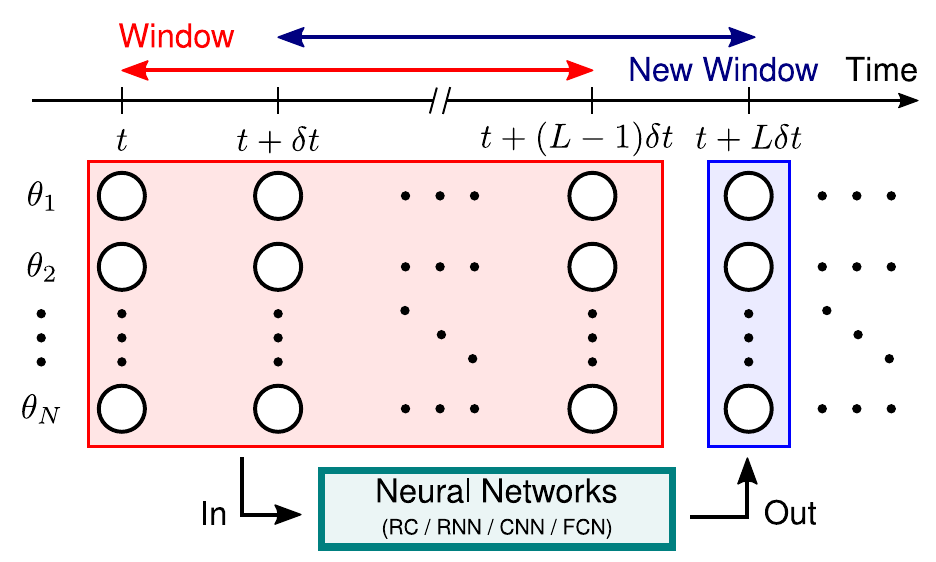}
\caption{Schematic of learning processes by NNs for prediction of phase evolution. Using the feedback process for a given length of the time window, $L$, the inputs for all models are determined from the phase evolution data $\{ \theta_i (t) \}$ for all oscillators. 
\label{fig:fig3}
}
\end{figure}

\begin{figure}[!h]
	\centering
	\includegraphics[width=0.98\linewidth]{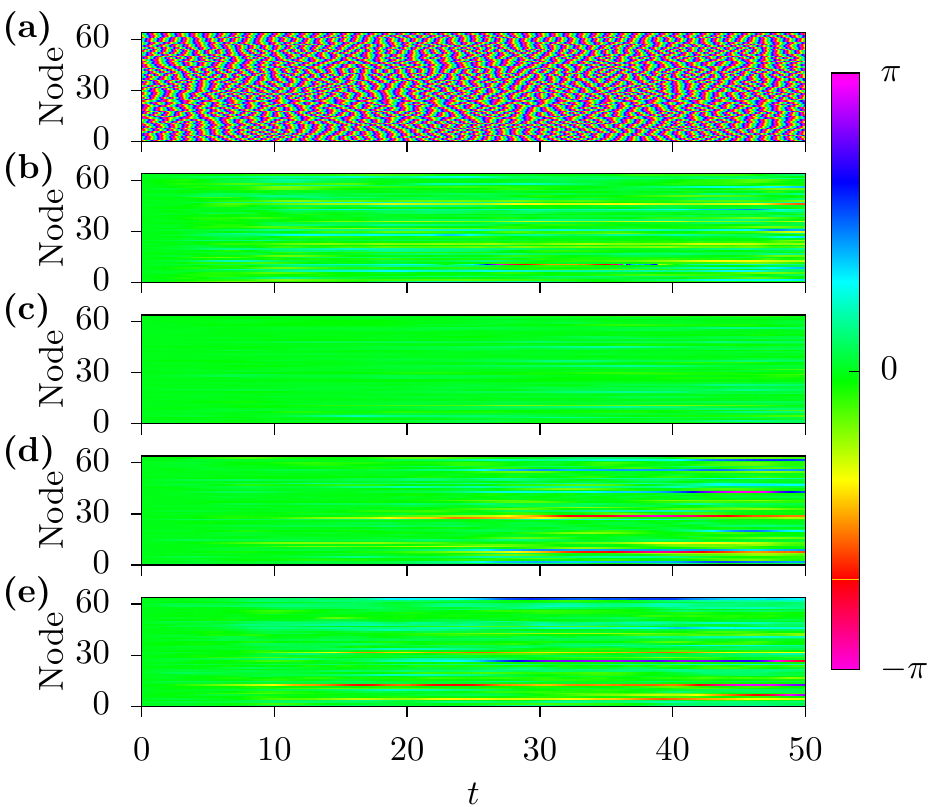}
	\caption{Prediction for the phase dynamics of the Kuramoto oscillators using four ML methods: (a) Evolution of $\{ \theta_i (t) \}$. (b)--(e) Difference between the actual data and predicted solution obtained using (b) RC, (c) classical RNN, (d) CNN, and (e) FCN.
		\label{fig:fig4}
	}
\end{figure}

As an input for this study, a time series of phase $\theta_i (t)$ is generated for each oscillator. We implement the fourth-order Runge-Kutta method with a discrete time step $\delta t = 0.005$ up to a total of $1.95 \times 10^5$ time steps for generating these datasets. To examine the predictability of the behaviors of all oscillators, for RC, we assume $8 \times 10^4$ steps of $\{\cos \theta_i (t), \sin \theta_i (t)\}$ as the washout period and the subsequent $10^5$ steps as training data to produce phases for the last $10^4$ time steps by feeding the output data back to the reservoir. For other models, considering $1.8 \times 10^5$ steps for training datasets, the subsequent $10^4$ time steps of phases are produced as an output by the NN for comparison with the exact dynamics of $\theta_i (t)$. 
A detailed description of these methods is illustrated in Fig.~\ref{fig:fig3}. For RC, we set the time length $L=1$ to predict the future phase dynamics, while $L=200$ is given as the input of the other models.

Fig.~\ref{fig:fig4} depicts the prediction for phase dynamics using the four ML models. As illustrated in Figs.~\ref{fig:fig4} (b)--(e), all types of NNs produce accurate phase dynamics data up to time $t=10$ (2000 time steps). Thus, although the Kuramoto model defined in Eq.~\eqref{eq:KM} exhibits nonlinearity and chaotic behavior for oscillators, ML approaches can be applied to learn the behavior of the phase dynamics and predict future dynamic patterns. In particular, the classical RNN method is the most beneficial among the four methods. 

The Kuramoto equation can be transformed into $\dot s_i =\omega_i c_i + c_i\sum_j (s_i c_j - c_i s_j)$, where $s_i$ and $c_i$ denote $\sin \theta_i$ and $\cos \theta_i$, respectively. This has the form of $s_i(t)=F(s_i(t-1), s_i(t-2),...., c_i(t-1), c_i(t-2), ....)$. Since feedback connections in RNNs are appropriate for predicting the future behavior of such nonlinear dynamical form of system, we can expect that it works well for not only nonlinear AR (autoregressive) model, but also for other RNN-based architectures such as LSTM (long short-term memory) and RC. As the LSTM is one of the most commonly used RNN for processing time series that can retain information for a long period of time, we adopted the LSTM in this paper.


\begin{thebibliography}{45}%
	\makeatletter
	\providecommand \@ifxundefined [1]{%
	 \@ifx{#1\undefined}
	}%
	\providecommand \@ifnum [1]{%
	 \ifnum #1\expandafter \@firstoftwo
	 \else \expandafter \@secondoftwo
	 \fi
	}%
	\providecommand \@ifx [1]{%
	 \ifx #1\expandafter \@firstoftwo
	 \else \expandafter \@secondoftwo
	 \fi
	}%
	\providecommand \natexlab [1]{#1}%
	\providecommand \enquote  [1]{``#1''}%
	\providecommand \bibnamefont  [1]{#1}%
	\providecommand \bibfnamefont [1]{#1}%
	\providecommand \citenamefont [1]{#1}%
	\providecommand \href@noop [0]{\@secondoftwo}%
	\providecommand \href [0]{\begingroup \@sanitize@url \@href}%
	\providecommand \@href[1]{\@@startlink{#1}\@@href}%
	\providecommand \@@href[1]{\endgroup#1\@@endlink}%
	\providecommand \@sanitize@url [0]{\catcode `\\12\catcode `\$12\catcode
	  `\&12\catcode `\#12\catcode `\^12\catcode `\_12\catcode `\%12\relax}%
	\providecommand \@@startlink[1]{}%
	\providecommand \@@endlink[0]{}%
	\providecommand \url  [0]{\begingroup\@sanitize@url \@url }%
	\providecommand \@url [1]{\endgroup\@href {#1}{\urlprefix }}%
	\providecommand \urlprefix  [0]{URL }%
	\providecommand \Eprint [0]{\href }%
	\providecommand \doibase [0]{http://dx.doi.org/}%
	\providecommand \selectlanguage [0]{\@gobble}%
	\providecommand \bibinfo  [0]{\@secondoftwo}%
	\providecommand \bibfield  [0]{\@secondoftwo}%
	\providecommand \translation [1]{[#1]}%
	\providecommand \BibitemOpen [0]{}%
	\providecommand \bibitemStop [0]{}%
	\providecommand \bibitemNoStop [0]{.\EOS\space}%
	\providecommand \EOS [0]{\spacefactor3000\relax}%
	\providecommand \BibitemShut  [1]{\csname bibitem#1\endcsname}%
	\let\auto@bib@innerbib\@empty
	\bibitem [{\citenamefont {Maass}\ \emph {et~al.}(2002)\citenamefont {Maass},
	  \citenamefont {Natschläger},\ and\ \citenamefont {Markram}}]{maass2002real}%
	  \BibitemOpen
	  \bibfield  {author} {\bibinfo {author} {\bibfnamefont {W.}~\bibnamefont
	  {Maass}}, \bibinfo {author} {\bibfnamefont {T.}~\bibnamefont {Natschläger}},
	  \ and\ \bibinfo {author} {\bibfnamefont {H.}~\bibnamefont {Markram}},\ }\href
	  {\doibase 10.1162/089976602760407955} {\bibfield  {journal} {\bibinfo
	  {journal} {Neur. Comp.}\ }\textbf {\bibinfo {volume} {14}},\ \bibinfo {pages}
	  {2531} (\bibinfo {year} {2002})}\BibitemShut {NoStop}%
	\bibitem [{\citenamefont {Jaeger}\ and\ \citenamefont
	  {Haas}(2004)}]{Jaeger2004}%
	  \BibitemOpen
	  \bibfield  {author} {\bibinfo {author} {\bibfnamefont {H.}~\bibnamefont
	  {Jaeger}}\ and\ \bibinfo {author} {\bibfnamefont {H.}~\bibnamefont {Haas}},\
	  }\href {\doibase 10.1126/science.1091277} {\bibfield  {journal} {\bibinfo
	  {journal} {Science}\ }\textbf {\bibinfo {volume} {304}},\ \bibinfo {pages}
	  {78} (\bibinfo {year} {2004})}\BibitemShut {NoStop}%
	\bibitem [{\citenamefont {Lukoševičius}\ and\ \citenamefont
	  {Jaeger}(2009)}]{Lukosevicius2009}%
	  \BibitemOpen
	  \bibfield  {author} {\bibinfo {author} {\bibfnamefont {M.}~\bibnamefont
	  {Lukoševičius}}\ and\ \bibinfo {author} {\bibfnamefont {H.}~\bibnamefont
	  {Jaeger}},\ }\href {\doibase https://doi.org/10.1016/j.cosrev.2009.03.005}
	  {\bibfield  {journal} {\bibinfo  {journal} {Comput. Sci. Rev.}\ }\textbf
	  {\bibinfo {volume} {3}},\ \bibinfo {pages} {127 } (\bibinfo {year}
	  {2009})}\BibitemShut {NoStop}%
	\bibitem [{\citenamefont {Lu}\ \emph {et~al.}(2017)\citenamefont {Lu},
	  \citenamefont {Pathak}, \citenamefont {Hunt}, \citenamefont {Girvan},
	  \citenamefont {Brockett},\ and\ \citenamefont {Ott}}]{Lu2017}%
	  \BibitemOpen
	  \bibfield  {author} {\bibinfo {author} {\bibfnamefont {Z.}~\bibnamefont
	  {Lu}}, \bibinfo {author} {\bibfnamefont {J.}~\bibnamefont {Pathak}}, \bibinfo
	  {author} {\bibfnamefont {B.}~\bibnamefont {Hunt}}, \bibinfo {author}
	  {\bibfnamefont {M.}~\bibnamefont {Girvan}}, \bibinfo {author} {\bibfnamefont
	  {R.}~\bibnamefont {Brockett}}, \ and\ \bibinfo {author} {\bibfnamefont
	  {E.}~\bibnamefont {Ott}},\ }\href {\doibase 10.1063/1.4979665} {\bibfield
	  {journal} {\bibinfo  {journal} {Chaos}\ }\textbf {\bibinfo {volume} {27}},\
	  \bibinfo {pages} {041102} (\bibinfo {year} {2017})}\BibitemShut {NoStop}%
	\bibitem [{\citenamefont {Carroll}(2018)}]{Carroll2018}%
	  \BibitemOpen
	  \bibfield  {author} {\bibinfo {author} {\bibfnamefont {T.~L.}\ \bibnamefont
	  {Carroll}},\ }\href {\doibase 10.1103/PhysRevE.98.052209} {\bibfield
	  {journal} {\bibinfo  {journal} {Phys. Rev. E}\ }\textbf {\bibinfo {volume}
	  {98}},\ \bibinfo {pages} {052209} (\bibinfo {year} {2018})}\BibitemShut
	  {NoStop}%
	\bibitem [{\citenamefont {Lu}\ \emph {et~al.}(2018)\citenamefont {Lu},
	  \citenamefont {Hunt},\ and\ \citenamefont {Ott}}]{Lu2018}%
	  \BibitemOpen
	  \bibfield  {author} {\bibinfo {author} {\bibfnamefont {Z.}~\bibnamefont
	  {Lu}}, \bibinfo {author} {\bibfnamefont {B.~R.}\ \bibnamefont {Hunt}}, \ and\
	  \bibinfo {author} {\bibfnamefont {E.}~\bibnamefont {Ott}},\ }\href {\doibase
	  10.1063/1.5039508} {\bibfield  {journal} {\bibinfo  {journal} {Chaos}\
	  }\textbf {\bibinfo {volume} {28}},\ \bibinfo {pages} {061104} (\bibinfo
	  {year} {2018})}\BibitemShut {NoStop}%
	\bibitem [{\citenamefont {Pathak}\ \emph {et~al.}(2017)\citenamefont {Pathak},
	  \citenamefont {Lu}, \citenamefont {Hunt}, \citenamefont {Girvan},\ and\
	  \citenamefont {Ott}}]{Pathak2017}%
	  \BibitemOpen
	  \bibfield  {author} {\bibinfo {author} {\bibfnamefont {J.}~\bibnamefont
	  {Pathak}}, \bibinfo {author} {\bibfnamefont {Z.}~\bibnamefont {Lu}}, \bibinfo
	  {author} {\bibfnamefont {B.~R.}\ \bibnamefont {Hunt}}, \bibinfo {author}
	  {\bibfnamefont {M.}~\bibnamefont {Girvan}}, \ and\ \bibinfo {author}
	  {\bibfnamefont {E.}~\bibnamefont {Ott}},\ }\href {\doibase 10.1063/1.5010300}
	  {\bibfield  {journal} {\bibinfo  {journal} {Chaos}\ }\textbf {\bibinfo
	  {volume} {27}},\ \bibinfo {pages} {121102} (\bibinfo {year}
	  {2017})}\BibitemShut {NoStop}%
	\bibitem [{\citenamefont {Pathak}\ \emph {et~al.}(2018)\citenamefont {Pathak},
	  \citenamefont {Hunt}, \citenamefont {Girvan}, \citenamefont {Lu},\ and\
	  \citenamefont {Ott}}]{Pathak2018}%
	  \BibitemOpen
	  \bibfield  {author} {\bibinfo {author} {\bibfnamefont {J.}~\bibnamefont
	  {Pathak}}, \bibinfo {author} {\bibfnamefont {B.}~\bibnamefont {Hunt}},
	  \bibinfo {author} {\bibfnamefont {M.}~\bibnamefont {Girvan}}, \bibinfo
	  {author} {\bibfnamefont {Z.}~\bibnamefont {Lu}}, \ and\ \bibinfo {author}
	  {\bibfnamefont {E.}~\bibnamefont {Ott}},\ }\href {\doibase
	  10.1103/PhysRevLett.120.024102} {\bibfield  {journal} {\bibinfo  {journal}
	  {Phys. Rev. Lett.}\ }\textbf {\bibinfo {volume} {120}},\ \bibinfo {pages}
	  {024102} (\bibinfo {year} {2018})}\BibitemShut {NoStop}%
	\bibitem [{\citenamefont {Weng}\ \emph {et~al.}(2019)\citenamefont {Weng},
	  \citenamefont {Yang}, \citenamefont {Gu}, \citenamefont {Zhang},\ and\
	  \citenamefont {Small}}]{Weng2019}%
	  \BibitemOpen
	  \bibfield  {author} {\bibinfo {author} {\bibfnamefont {T.}~\bibnamefont
	  {Weng}}, \bibinfo {author} {\bibfnamefont {H.}~\bibnamefont {Yang}}, \bibinfo
	  {author} {\bibfnamefont {C.}~\bibnamefont {Gu}}, \bibinfo {author}
	  {\bibfnamefont {J.}~\bibnamefont {Zhang}}, \ and\ \bibinfo {author}
	  {\bibfnamefont {M.}~\bibnamefont {Small}},\ }\href {\doibase
	  10.1103/PhysRevE.99.042203} {\bibfield  {journal} {\bibinfo  {journal} {Phys.
	  Rev. E}\ }\textbf {\bibinfo {volume} {99}},\ \bibinfo {pages} {042203}
	  (\bibinfo {year} {2019})}\BibitemShut {NoStop}%
	\bibitem [{\citenamefont {Jiang}\ and\ \citenamefont {Lai}(2019)}]{Jiang2019}%
	  \BibitemOpen
	  \bibfield  {author} {\bibinfo {author} {\bibfnamefont {J.}~\bibnamefont
	  {Jiang}}\ and\ \bibinfo {author} {\bibfnamefont {Y.-C.}\ \bibnamefont
	  {Lai}},\ }\href {\doibase 10.1103/PhysRevResearch.1.033056} {\bibfield
	  {journal} {\bibinfo  {journal} {Phys. Rev. Res.}\ }\textbf {\bibinfo {volume}
	  {1}},\ \bibinfo {pages} {033056} (\bibinfo {year} {2019})}\BibitemShut
	  {NoStop}%
	\bibitem [{\citenamefont {Fan}\ \emph {et~al.}(2020)\citenamefont {Fan},
	  \citenamefont {Jiang}, \citenamefont {Zhang}, \citenamefont {Wang},\ and\
	  \citenamefont {Lai}}]{Fan2020}%
	  \BibitemOpen
	  \bibfield  {author} {\bibinfo {author} {\bibfnamefont {H.}~\bibnamefont
	  {Fan}}, \bibinfo {author} {\bibfnamefont {J.}~\bibnamefont {Jiang}}, \bibinfo
	  {author} {\bibfnamefont {C.}~\bibnamefont {Zhang}}, \bibinfo {author}
	  {\bibfnamefont {X.}~\bibnamefont {Wang}}, \ and\ \bibinfo {author}
	  {\bibfnamefont {Y.-C.}\ \bibnamefont {Lai}},\ }\href {\doibase
	  10.1103/PhysRevResearch.2.012080} {\bibfield  {journal} {\bibinfo  {journal}
	  {Phys. Rev. Res.}\ }\textbf {\bibinfo {volume} {2}},\ \bibinfo {pages}
	  {012080} (\bibinfo {year} {2020})}\BibitemShut {NoStop}%
	\bibitem [{\citenamefont {Zhang}\ \emph {et~al.}(2020)\citenamefont {Zhang},
	  \citenamefont {Jiang}, \citenamefont {Qu},\ and\ \citenamefont
	  {Lai}}]{Lai2020}%
	  \BibitemOpen
	  \bibfield  {author} {\bibinfo {author} {\bibfnamefont {C.}~\bibnamefont
	  {Zhang}}, \bibinfo {author} {\bibfnamefont {J.}~\bibnamefont {Jiang}},
	  \bibinfo {author} {\bibfnamefont {S.-X.}\ \bibnamefont {Qu}}, \ and\ \bibinfo
	  {author} {\bibfnamefont {Y.-C.}\ \bibnamefont {Lai}},\ }\href {\doibase
	  10.1063/5.0006304} {\bibfield  {journal} {\bibinfo  {journal} {Chaos}\
	  }\textbf {\bibinfo {volume} {30}},\ \bibinfo {pages} {083114} (\bibinfo
	  {year} {2020})}\BibitemShut {NoStop}%
	\bibitem [{\citenamefont {Nitzan}\ \emph {et~al.}(2017)\citenamefont {Nitzan},
	  \citenamefont {Casadiego},\ and\ \citenamefont {Timme}}]{Nitzan2017}%
	  \BibitemOpen
	  \bibfield  {author} {\bibinfo {author} {\bibfnamefont {M.}~\bibnamefont
	  {Nitzan}}, \bibinfo {author} {\bibfnamefont {J.}~\bibnamefont {Casadiego}}, \
	  and\ \bibinfo {author} {\bibfnamefont {M.}~\bibnamefont {Timme}},\ }\href
	  {https://advances.sciencemag.org/content/3/2/e1600396} {\bibfield  {journal}
	  {\bibinfo  {journal} {Sci. Adv.}\ }\textbf {\bibinfo {volume} {3}} (\bibinfo
	  {year} {2017})}\BibitemShut {NoStop}%
	\bibitem [{\citenamefont {Wang}\ \emph {et~al.}(2016)\citenamefont {Wang},
	  \citenamefont {Lai},\ and\ \citenamefont {Grebogi}}]{wang2016}%
	  \BibitemOpen
	  \bibfield  {author} {\bibinfo {author} {\bibfnamefont {W.-X.}\ \bibnamefont
	  {Wang}}, \bibinfo {author} {\bibfnamefont {Y.-C.}\ \bibnamefont {Lai}}, \
	  and\ \bibinfo {author} {\bibfnamefont {C.}~\bibnamefont {Grebogi}},\ }\href
	  {\doibase https://doi.org/10.1016/j.physrep.2016.06.004} {\bibfield
	  {journal} {\bibinfo  {journal} {Phys. Rep.}\ }\textbf {\bibinfo {volume}
	  {644}},\ \bibinfo {pages} {1 } (\bibinfo {year} {2016})}\BibitemShut
	  {NoStop}%
	\bibitem [{\citenamefont {Eroglu}\ \emph {et~al.}(2020)\citenamefont {Eroglu},
	  \citenamefont {Tanzi}, \citenamefont {van Strien},\ and\ \citenamefont
	  {Pereira}}]{eroglu2020}%
	  \BibitemOpen
	  \bibfield  {author} {\bibinfo {author} {\bibfnamefont {D.}~\bibnamefont
	  {Eroglu}}, \bibinfo {author} {\bibfnamefont {M.}~\bibnamefont {Tanzi}},
	  \bibinfo {author} {\bibfnamefont {S.}~\bibnamefont {van Strien}}, \ and\
	  \bibinfo {author} {\bibfnamefont {T.}~\bibnamefont {Pereira}},\ }\href
	  {\doibase 10.1103/PhysRevX.10.021047} {\bibfield  {journal} {\bibinfo
	  {journal} {Phys. Rev. X}\ }\textbf {\bibinfo {volume} {10}},\ \bibinfo
	  {pages} {021047} (\bibinfo {year} {2020})}\BibitemShut {NoStop}%
	\bibitem [{\citenamefont {Mormann}\ \emph {et~al.}(2005)\citenamefont
	  {Mormann}, \citenamefont {Kreuz}, \citenamefont {Rieke}, \citenamefont
	  {Andrzejak}, \citenamefont {Kraskov}, \citenamefont {David}, \citenamefont
	  {Elger},\ and\ \citenamefont {Lehnertz}}]{mormann2005predictability}%
	  \BibitemOpen
	  \bibfield  {author} {\bibinfo {author} {\bibfnamefont {F.}~\bibnamefont
	  {Mormann}}, \bibinfo {author} {\bibfnamefont {T.}~\bibnamefont {Kreuz}},
	  \bibinfo {author} {\bibfnamefont {C.}~\bibnamefont {Rieke}}, \bibinfo
	  {author} {\bibfnamefont {R.~G.}\ \bibnamefont {Andrzejak}}, \bibinfo {author}
	  {\bibfnamefont {A.}~\bibnamefont {Kraskov}}, \bibinfo {author} {\bibfnamefont
	  {P.}~\bibnamefont {David}}, \bibinfo {author} {\bibfnamefont {C.~E.}\
	  \bibnamefont {Elger}}, \ and\ \bibinfo {author} {\bibfnamefont
	  {K.}~\bibnamefont {Lehnertz}},\ }\href {\doibase
	  https://doi.org/10.1016/j.clinph.2004.08.025} {\bibfield  {journal} {\bibinfo
	   {journal} {Clin. Neurophysiol.}\ }\textbf {\bibinfo {volume} {116}},\
	  \bibinfo {pages} {569 } (\bibinfo {year} {2005})}\BibitemShut {NoStop}%
	\bibitem [{\citenamefont {Mirowski}\ \emph {et~al.}(2009)\citenamefont
	  {Mirowski}, \citenamefont {Madhavan}, \citenamefont {LeCun},\ and\
	  \citenamefont {Kuzniecky}}]{mirowski2009classification}%
	  \BibitemOpen
	  \bibfield  {author} {\bibinfo {author} {\bibfnamefont {P.}~\bibnamefont
	  {Mirowski}}, \bibinfo {author} {\bibfnamefont {D.}~\bibnamefont {Madhavan}},
	  \bibinfo {author} {\bibfnamefont {Y.}~\bibnamefont {LeCun}}, \ and\ \bibinfo
	  {author} {\bibfnamefont {R.}~\bibnamefont {Kuzniecky}},\ }\href {\doibase
	  https://doi.org/10.1016/j.clinph.2009.09.002} {\bibfield  {journal} {\bibinfo
	   {journal} {Clin. Neurophysiol.}\ }\textbf {\bibinfo {volume} {120}},\
	  \bibinfo {pages} {1927 } (\bibinfo {year} {2009})}\BibitemShut {NoStop}%
	\bibitem [{\citenamefont {Chandaka}\ \emph {et~al.}(2009)\citenamefont
	  {Chandaka}, \citenamefont {Chatterjee},\ and\ \citenamefont
	  {Munshi}}]{chandaka2009cross}%
	  \BibitemOpen
	  \bibfield  {author} {\bibinfo {author} {\bibfnamefont {S.}~\bibnamefont
	  {Chandaka}}, \bibinfo {author} {\bibfnamefont {A.}~\bibnamefont
	  {Chatterjee}}, \ and\ \bibinfo {author} {\bibfnamefont {S.}~\bibnamefont
	  {Munshi}},\ }\href {\doibase https://doi.org/10.1016/j.eswa.2007.11.017}
	  {\bibfield  {journal} {\bibinfo  {journal} {Expert Syst. Appl.}\ }\textbf
	  {\bibinfo {volume} {36}},\ \bibinfo {pages} {1329 } (\bibinfo {year}
	  {2009})}\BibitemShut {NoStop}%
	\bibitem [{\citenamefont {Williamson}\ \emph {et~al.}(2012)\citenamefont
	  {Williamson}, \citenamefont {Bliss}, \citenamefont {Browne},\ and\
	  \citenamefont {Narayanan}}]{williamson2012seizure}%
	  \BibitemOpen
	  \bibfield  {author} {\bibinfo {author} {\bibfnamefont {J.~R.}\ \bibnamefont
	  {Williamson}}, \bibinfo {author} {\bibfnamefont {D.~W.}\ \bibnamefont
	  {Bliss}}, \bibinfo {author} {\bibfnamefont {D.~W.}\ \bibnamefont {Browne}}, \
	  and\ \bibinfo {author} {\bibfnamefont {J.~T.}\ \bibnamefont {Narayanan}},\
	  }\href {\doibase https://doi.org/10.1016/j.yebeh.2012.07.007} {\bibfield
	  {journal} {\bibinfo  {journal} {Epilepsy Behav.}\ }\textbf {\bibinfo {volume}
	  {25}},\ \bibinfo {pages} {230 } (\bibinfo {year} {2012})}\BibitemShut
	  {NoStop}%
	\bibitem [{\citenamefont {Bohrdt}\ \emph {et~al.}(2019)\citenamefont {Bohrdt},
	  \citenamefont {Chiu}, \citenamefont {Ji}, \citenamefont {Xu}, \citenamefont
	  {Greif}, \citenamefont {Greiner}, \citenamefont {Demler}, \citenamefont
	  {Grusdt},\ and\ \citenamefont {Knap}}]{Bohrdt2019}%
	  \BibitemOpen
	  \bibfield  {author} {\bibinfo {author} {\bibfnamefont {A.}~\bibnamefont
	  {Bohrdt}}, \bibinfo {author} {\bibfnamefont {C.~S.}\ \bibnamefont {Chiu}},
	  \bibinfo {author} {\bibfnamefont {G.}~\bibnamefont {Ji}}, \bibinfo {author}
	  {\bibfnamefont {M.}~\bibnamefont {Xu}}, \bibinfo {author} {\bibfnamefont
	  {D.}~\bibnamefont {Greif}}, \bibinfo {author} {\bibfnamefont
	  {M.}~\bibnamefont {Greiner}}, \bibinfo {author} {\bibfnamefont
	  {E.}~\bibnamefont {Demler}}, \bibinfo {author} {\bibfnamefont
	  {F.}~\bibnamefont {Grusdt}}, \ and\ \bibinfo {author} {\bibfnamefont
	  {M.}~\bibnamefont {Knap}},\ }\href@noop {} {\bibfield  {journal} {\bibinfo
	  {journal} {Nat. Phys.}\ }\textbf {\bibinfo {volume} {15}},\ \bibinfo {pages}
	  {921} (\bibinfo {year} {2019})}\BibitemShut {NoStop}%
	\bibitem [{\citenamefont {Zhang}\ \emph {et~al.}(2019)\citenamefont {Zhang},
	  \citenamefont {Liu},\ and\ \citenamefont {Wei}}]{Zhang2019}%
	  \BibitemOpen
	  \bibfield  {author} {\bibinfo {author} {\bibfnamefont {W.}~\bibnamefont
	  {Zhang}}, \bibinfo {author} {\bibfnamefont {J.}~\bibnamefont {Liu}}, \ and\
	  \bibinfo {author} {\bibfnamefont {T.-C.}\ \bibnamefont {Wei}},\ }\href
	  {\doibase 10.1103/PhysRevE.99.032142} {\bibfield  {journal} {\bibinfo
	  {journal} {Phys. Rev. E}\ }\textbf {\bibinfo {volume} {99}},\ \bibinfo
	  {pages} {032142} (\bibinfo {year} {2019})}\BibitemShut {NoStop}%
	\bibitem [{\citenamefont {Carrasquilla}\ and\ \citenamefont
	  {Melko}(2017)}]{Carrasquilla2017}%
	  \BibitemOpen
	  \bibfield  {author} {\bibinfo {author} {\bibfnamefont {J.}~\bibnamefont
	  {Carrasquilla}}\ and\ \bibinfo {author} {\bibfnamefont {R.~G.}\ \bibnamefont
	  {Melko}},\ }\href@noop {} {\bibfield  {journal} {\bibinfo  {journal} {Nat.
	  Phys.}\ }\textbf {\bibinfo {volume} {13}},\ \bibinfo {pages} {431} (\bibinfo
	  {year} {2017})}\BibitemShut {NoStop}%
	\bibitem [{\citenamefont {Venderley}\ \emph {et~al.}(2018)\citenamefont
	  {Venderley}, \citenamefont {Khemani},\ and\ \citenamefont
	  {Kim}}]{Venderley2018}%
	  \BibitemOpen
	  \bibfield  {author} {\bibinfo {author} {\bibfnamefont {J.}~\bibnamefont
	  {Venderley}}, \bibinfo {author} {\bibfnamefont {V.}~\bibnamefont {Khemani}},
	  \ and\ \bibinfo {author} {\bibfnamefont {E.-A.}\ \bibnamefont {Kim}},\ }\href
	  {\doibase 10.1103/PhysRevLett.120.257204} {\bibfield  {journal} {\bibinfo
	  {journal} {Phys. Rev. Lett.}\ }\textbf {\bibinfo {volume} {120}},\ \bibinfo
	  {pages} {257204} (\bibinfo {year} {2018})}\BibitemShut {NoStop}%
	\bibitem [{\citenamefont {Beach}\ \emph {et~al.}(2018)\citenamefont {Beach},
	  \citenamefont {Golubeva},\ and\ \citenamefont {Melko}}]{Beach2018}%
	  \BibitemOpen
	  \bibfield  {author} {\bibinfo {author} {\bibfnamefont {M.~J.~S.}\
	  \bibnamefont {Beach}}, \bibinfo {author} {\bibfnamefont {A.}~\bibnamefont
	  {Golubeva}}, \ and\ \bibinfo {author} {\bibfnamefont {R.~G.}\ \bibnamefont
	  {Melko}},\ }\href {\doibase 10.1103/PhysRevB.97.045207} {\bibfield  {journal}
	  {\bibinfo  {journal} {Phys. Rev. B}\ }\textbf {\bibinfo {volume} {97}},\
	  \bibinfo {pages} {045207} (\bibinfo {year} {2018})}\BibitemShut {NoStop}%
	\bibitem [{\citenamefont {Ni}\ \emph {et~al.}(2019)\citenamefont {Ni},
	  \citenamefont {Tang}, \citenamefont {Liu},\ and\ \citenamefont
	  {Lai}}]{Ni2019}%
	  \BibitemOpen
	  \bibfield  {author} {\bibinfo {author} {\bibfnamefont {Q.}~\bibnamefont
	  {Ni}}, \bibinfo {author} {\bibfnamefont {M.}~\bibnamefont {Tang}}, \bibinfo
	  {author} {\bibfnamefont {Y.}~\bibnamefont {Liu}}, \ and\ \bibinfo {author}
	  {\bibfnamefont {Y.-C.}\ \bibnamefont {Lai}},\ }\href {\doibase
	  10.1103/PhysRevE.100.052312} {\bibfield  {journal} {\bibinfo  {journal}
	  {Phys. Rev. E}\ }\textbf {\bibinfo {volume} {100}},\ \bibinfo {pages}
	  {052312} (\bibinfo {year} {2019})}\BibitemShut {NoStop}%
	\bibitem [{\citenamefont {Broecker}\ \emph {et~al.}(2017)\citenamefont
	  {Broecker}, \citenamefont {Carrasquilla}, \citenamefont {Melko},\ and\
	  \citenamefont {Trebst}}]{Broecker2017}%
	  \BibitemOpen
	  \bibfield  {author} {\bibinfo {author} {\bibfnamefont {P.}~\bibnamefont
	  {Broecker}}, \bibinfo {author} {\bibfnamefont {J.}~\bibnamefont
	  {Carrasquilla}}, \bibinfo {author} {\bibfnamefont {R.~G.}\ \bibnamefont
	  {Melko}}, \ and\ \bibinfo {author} {\bibfnamefont {S.}~\bibnamefont
	  {Trebst}},\ }\href@noop {} {\bibfield  {journal} {\bibinfo  {journal} {Sci.
	  Rep.}\ }\textbf {\bibinfo {volume} {7}},\ \bibinfo {pages} {1} (\bibinfo
	  {year} {2017})}\BibitemShut {NoStop}%
	\bibitem [{\citenamefont {Kuramoto}(1975)}]{Kuramoto1975}%
	  \BibitemOpen
	  \bibfield  {author} {\bibinfo {author} {\bibfnamefont {Y.}~\bibnamefont
	  {Kuramoto}},\ }\href {\doibase 10.1007/BFb0013365} {\emph {\bibinfo {title}
	  {International Symposium on Mathematical Problems in Theoretical Physics}}}\
	  (\bibinfo  {publisher} {Springer-Verlag},\ \bibinfo {address}
	  {Berlin/Heidelberg},\ \bibinfo {year} {1975})\ pp.\ \bibinfo {pages}
	  {420--422}\BibitemShut {NoStop}%
	\bibitem [{\citenamefont {Kuramoto}(1984)}]{Kuramoto1984}%
	  \BibitemOpen
	  \bibfield  {author} {\bibinfo {author} {\bibfnamefont {Y.}~\bibnamefont
	  {Kuramoto}},\ }\href {\doibase 10.1007/978-3-642-69689-3} {\emph {\bibinfo
	  {title} {{Chemical Oscillations, Waves, and Turbulence}}}},\ \bibinfo
	  {series} {Springer Series in Synergetics}, Vol.~\bibinfo {volume} {19}\
	  (\bibinfo  {publisher} {Springer Berlin Heidelberg},\ \bibinfo {address}
	  {Berlin, Heidelberg},\ \bibinfo {year} {1984})\BibitemShut {NoStop}%
	\bibitem [{\citenamefont {Martens}\ \emph {et~al.}(2009)\citenamefont
	  {Martens}, \citenamefont {Barreto}, \citenamefont {Strogatz}, \citenamefont
	  {Ott}, \citenamefont {So},\ and\ \citenamefont
	  {Antonsen}}]{martens2009exact}%
	  \BibitemOpen
	  \bibfield  {author} {\bibinfo {author} {\bibfnamefont {E.~A.}\ \bibnamefont
	  {Martens}}, \bibinfo {author} {\bibfnamefont {E.}~\bibnamefont {Barreto}},
	  \bibinfo {author} {\bibfnamefont {S.~H.}\ \bibnamefont {Strogatz}}, \bibinfo
	  {author} {\bibfnamefont {E.}~\bibnamefont {Ott}}, \bibinfo {author}
	  {\bibfnamefont {P.}~\bibnamefont {So}}, \ and\ \bibinfo {author}
	  {\bibfnamefont {T.~M.}\ \bibnamefont {Antonsen}},\ }\href {\doibase
	  10.1103/PhysRevE.79.026204} {\bibfield  {journal} {\bibinfo  {journal} {Phys.
	  Rev. E}\ }\textbf {\bibinfo {volume} {79}},\ \bibinfo {pages} {026204}
	  (\bibinfo {year} {2009})}\BibitemShut {NoStop}%
	\bibitem [{\citenamefont {Paz\'o}\ and\ \citenamefont
	  {Montbri\'o}(2009)}]{pazo2009existence}%
	  \BibitemOpen
	  \bibfield  {author} {\bibinfo {author} {\bibfnamefont {D.}~\bibnamefont
	  {Paz\'o}}\ and\ \bibinfo {author} {\bibfnamefont {E.}~\bibnamefont
	  {Montbri\'o}},\ }\href {\doibase 10.1103/PhysRevE.80.046215} {\bibfield
	  {journal} {\bibinfo  {journal} {Phys. Rev. E}\ }\textbf {\bibinfo {volume}
	  {80}},\ \bibinfo {pages} {046215} (\bibinfo {year} {2009})}\BibitemShut
	  {NoStop}%
	\bibitem [{\citenamefont {Skardal}(2018)}]{skardal2018low}%
	  \BibitemOpen
	  \bibfield  {author} {\bibinfo {author} {\bibfnamefont {P.~S.}\ \bibnamefont
	  {Skardal}},\ }\href {\doibase 10.1103/PhysRevE.98.022207} {\bibfield
	  {journal} {\bibinfo  {journal} {Phys. Rev. E}\ }\textbf {\bibinfo {volume}
	  {98}},\ \bibinfo {pages} {022207} (\bibinfo {year} {2018})}\BibitemShut
	  {NoStop}%
	\bibitem [{\citenamefont {Tang}(2011)}]{Tang2011}%
	  \BibitemOpen
	  \bibfield  {author} {\bibinfo {author} {\bibfnamefont {L.-H.}\ \bibnamefont
	  {Tang}},\ }\href {\doibase 10.1088/1742-5468/2011/01/p01034} {\bibfield
	  {journal} {\bibinfo  {journal} {J. Stat. Mech.: Theory Exp.}\ }\textbf
	  {\bibinfo {volume} {2011}},\ \bibinfo {pages} {P01034} (\bibinfo {year}
	  {2011})}\BibitemShut {NoStop}%
	\bibitem [{\citenamefont {Rodrigues}\ \emph {et~al.}(2016)\citenamefont
	  {Rodrigues}, \citenamefont {Peron}, \citenamefont {Ji},\ and\ \citenamefont
	  {Kurths}}]{Rodrigues2016}%
	  \BibitemOpen
	  \bibfield  {author} {\bibinfo {author} {\bibfnamefont {F.~A.}\ \bibnamefont
	  {Rodrigues}}, \bibinfo {author} {\bibfnamefont {T.~K.~D.}\ \bibnamefont
	  {Peron}}, \bibinfo {author} {\bibfnamefont {P.}~\bibnamefont {Ji}}, \ and\
	  \bibinfo {author} {\bibfnamefont {J.}~\bibnamefont {Kurths}},\ }\href
	  {\doibase https://doi.org/10.1016/j.physrep.2015.10.008} {\bibfield
	  {journal} {\bibinfo  {journal} {Phys. Rep.}\ }\textbf {\bibinfo {volume}
	  {610}},\ \bibinfo {pages} {1 } (\bibinfo {year} {2016})}\BibitemShut
	  {NoStop}%
	\bibitem [{\citenamefont {Paz\'o}(2005)}]{Pazo2005}%
	  \BibitemOpen
	  \bibfield  {author} {\bibinfo {author} {\bibfnamefont {D.}~\bibnamefont
	  {Paz\'o}},\ }\href {\doibase 10.1103/PhysRevE.72.046211} {\bibfield
	  {journal} {\bibinfo  {journal} {Phys. Rev. E}\ }\textbf {\bibinfo {volume}
	  {72}},\ \bibinfo {pages} {046211} (\bibinfo {year} {2005})}\BibitemShut
	  {NoStop}%
	\bibitem [{\citenamefont {Basnarkov}\ and\ \citenamefont
	  {Urumov}(2007)}]{Basnarkov2007}%
	  \BibitemOpen
	  \bibfield  {author} {\bibinfo {author} {\bibfnamefont {L.}~\bibnamefont
	  {Basnarkov}}\ and\ \bibinfo {author} {\bibfnamefont {V.}~\bibnamefont
	  {Urumov}},\ }\href {\doibase 10.1103/PhysRevE.76.057201} {\bibfield
	  {journal} {\bibinfo  {journal} {Phys. Rev. E}\ }\textbf {\bibinfo {volume}
	  {76}},\ \bibinfo {pages} {057201} (\bibinfo {year} {2007})}\BibitemShut
	  {NoStop}%
	\bibitem [{\citenamefont {Coutinho}\ \emph {et~al.}(2013)\citenamefont
	  {Coutinho}, \citenamefont {Goltsev}, \citenamefont {Dorogovtsev},\ and\
	  \citenamefont {Mendes}}]{Coutinho2013}%
	  \BibitemOpen
	  \bibfield  {author} {\bibinfo {author} {\bibfnamefont {B.~C.}\ \bibnamefont
	  {Coutinho}}, \bibinfo {author} {\bibfnamefont {A.~V.}\ \bibnamefont
	  {Goltsev}}, \bibinfo {author} {\bibfnamefont {S.~N.}\ \bibnamefont
	  {Dorogovtsev}}, \ and\ \bibinfo {author} {\bibfnamefont {J.~F.~F.}\
	  \bibnamefont {Mendes}},\ }\href {\doibase 10.1103/PhysRevE.87.032106}
	  {\bibfield  {journal} {\bibinfo  {journal} {Phys. Rev. E}\ }\textbf {\bibinfo
	  {volume} {87}},\ \bibinfo {pages} {032106} (\bibinfo {year}
	  {2013})}\BibitemShut {NoStop}%
	\bibitem [{\citenamefont {Song}\ \emph {et~al.}(2020)\citenamefont {Song},
	  \citenamefont {Um}, \citenamefont {Park},\ and\ \citenamefont
	  {Kahng}}]{Song2020}%
	  \BibitemOpen
	  \bibfield  {author} {\bibinfo {author} {\bibfnamefont {J.~U.}\ \bibnamefont
	  {Song}}, \bibinfo {author} {\bibfnamefont {J.}~\bibnamefont {Um}}, \bibinfo
	  {author} {\bibfnamefont {J.}~\bibnamefont {Park}}, \ and\ \bibinfo {author}
	  {\bibfnamefont {B.}~\bibnamefont {Kahng}},\ }\href {\doibase
	  10.1103/PhysRevE.101.052313} {\bibfield  {journal} {\bibinfo  {journal}
	  {Phys. Rev. E}\ }\textbf {\bibinfo {volume} {101}},\ \bibinfo {pages}
	  {052313} (\bibinfo {year} {2020})}\BibitemShut {NoStop}%
	\bibitem [{\citenamefont {Choi}\ \emph {et~al.}(2013)\citenamefont {Choi},
	  \citenamefont {Ha},\ and\ \citenamefont {Kahng}}]{Choi2013}%
	  \BibitemOpen
	  \bibfield  {author} {\bibinfo {author} {\bibfnamefont {C.}~\bibnamefont
	  {Choi}}, \bibinfo {author} {\bibfnamefont {M.}~\bibnamefont {Ha}}, \ and\
	  \bibinfo {author} {\bibfnamefont {B.}~\bibnamefont {Kahng}},\ }\href
	  {\doibase 10.1103/PhysRevE.88.032126} {\bibfield  {journal} {\bibinfo
	  {journal} {Phys. Rev. E}\ }\textbf {\bibinfo {volume} {88}},\ \bibinfo
	  {pages} {032126} (\bibinfo {year} {2013})}\BibitemShut {NoStop}%
	\bibitem [{\citenamefont {Yoon}\ \emph {et~al.}(2015)\citenamefont {Yoon},
	  \citenamefont {Sorbaro~Sindaci}, \citenamefont {Goltsev},\ and\ \citenamefont
	  {Mendes}}]{Yoon2015}%
	  \BibitemOpen
	  \bibfield  {author} {\bibinfo {author} {\bibfnamefont {S.}~\bibnamefont
	  {Yoon}}, \bibinfo {author} {\bibfnamefont {M.}~\bibnamefont
	  {Sorbaro~Sindaci}}, \bibinfo {author} {\bibfnamefont {A.~V.}\ \bibnamefont
	  {Goltsev}}, \ and\ \bibinfo {author} {\bibfnamefont {J.~F.~F.}\ \bibnamefont
	  {Mendes}},\ }\href {\doibase 10.1103/PhysRevE.91.032814} {\bibfield
	  {journal} {\bibinfo  {journal} {Phys. Rev. E}\ }\textbf {\bibinfo {volume}
	  {91}},\ \bibinfo {pages} {032814} (\bibinfo {year} {2015})}\BibitemShut
	  {NoStop}%
	\bibitem [{\citenamefont {Cho}\ \emph {et~al.}(2016)\citenamefont {Cho},
	  \citenamefont {Lee}, \citenamefont {Herrmann},\ and\ \citenamefont
	  {Kahng}}]{rer}%
	  \BibitemOpen
	  \bibfield  {author} {\bibinfo {author} {\bibfnamefont {Y.~S.}\ \bibnamefont
	  {Cho}}, \bibinfo {author} {\bibfnamefont {J.~S.}\ \bibnamefont {Lee}},
	  \bibinfo {author} {\bibfnamefont {H.~J.}\ \bibnamefont {Herrmann}}, \ and\
	  \bibinfo {author} {\bibfnamefont {B.}~\bibnamefont {Kahng}},\ }\href
	  {\doibase 10.1103/PhysRevLett.116.025701} {\bibfield  {journal} {\bibinfo
	  {journal} {Phys. Rev. Lett.}\ }\textbf {\bibinfo {volume} {116}},\ \bibinfo
	  {pages} {025701} (\bibinfo {year} {2016})}\BibitemShut {NoStop}%
	\bibitem [{\citenamefont {Goodfellow}\ \emph {et~al.}(2016)\citenamefont
	  {Goodfellow}, \citenamefont {Bengio},\ and\ \citenamefont
	  {Courville}}]{Goodfellow2016}%
	  \BibitemOpen
	  \bibfield  {author} {\bibinfo {author} {\bibfnamefont {I.}~\bibnamefont
	  {Goodfellow}}, \bibinfo {author} {\bibfnamefont {Y.}~\bibnamefont {Bengio}},
	  \ and\ \bibinfo {author} {\bibfnamefont {A.}~\bibnamefont {Courville}},\
	  }\href@noop {} {\emph {\bibinfo {title} {Deep Learning}}}\ (\bibinfo
	  {publisher} {MIT Press},\ \bibinfo {year} {2016})\BibitemShut {NoStop}%
	\bibitem [{\citenamefont {Oh}\ \emph {et~al.}(2005)\citenamefont {Oh},
	  \citenamefont {Rho}, \citenamefont {Hong},\ and\ \citenamefont {Kahng}}]{Oh}%
	  \BibitemOpen
	  \bibfield  {author} {\bibinfo {author} {\bibfnamefont {E.}~\bibnamefont
	  {Oh}}, \bibinfo {author} {\bibfnamefont {K.}~\bibnamefont {Rho}}, \bibinfo
	  {author} {\bibfnamefont {H.}~\bibnamefont {Hong}}, \ and\ \bibinfo {author}
	  {\bibfnamefont {B.}~\bibnamefont {Kahng}},\ }\href {\doibase
	  10.1103/PhysRevE.72.047101} {\bibfield  {journal} {\bibinfo  {journal} {Phys.
	  Rev. E}\ }\textbf {\bibinfo {volume} {72}},\ \bibinfo {pages} {047101}
	  (\bibinfo {year} {2005})}\BibitemShut {NoStop}%
	\bibitem [{\citenamefont {Arenas}\ \emph {et~al.}(2006)\citenamefont {Arenas},
	  \citenamefont {D\'{\i}az-Guilera},\ and\ \citenamefont
	  {P\'erez-Vicente}}]{Arenas2006}%
	  \BibitemOpen
	  \bibfield  {author} {\bibinfo {author} {\bibfnamefont {A.}~\bibnamefont
	  {Arenas}}, \bibinfo {author} {\bibfnamefont {A.}~\bibnamefont
	  {D\'{\i}az-Guilera}}, \ and\ \bibinfo {author} {\bibfnamefont {C.~J.}\
	  \bibnamefont {P\'erez-Vicente}},\ }\href {\doibase
	  10.1103/PhysRevLett.96.114102} {\bibfield  {journal} {\bibinfo  {journal}
	  {Phys. Rev. Lett.}\ }\textbf {\bibinfo {volume} {96}},\ \bibinfo {pages}
	  {114102} (\bibinfo {year} {2006})}\BibitemShut {NoStop}%
	\bibitem [{\citenamefont {Kim}\ \emph {et~al.}(2021)\citenamefont {Kim},
	  \citenamefont {Lu}, \citenamefont {Nozari}, \citenamefont {Pappas},\ and\
	  \citenamefont {Bassett}}]{Kim_2021}%
	  \BibitemOpen
	  \bibfield  {author} {\bibinfo {author} {\bibfnamefont {J.~Z.}\ \bibnamefont
	  {Kim}}, \bibinfo {author} {\bibfnamefont {Z.}~\bibnamefont {Lu}}, \bibinfo
	  {author} {\bibfnamefont {E.}~\bibnamefont {Nozari}}, \bibinfo {author}
	  {\bibfnamefont {G.~J.}\ \bibnamefont {Pappas}}, \ and\ \bibinfo {author}
	  {\bibfnamefont {D.~S.}\ \bibnamefont {Bassett}},\ }\href {\doibase
	  10.1038/s42256-021-00321-2} {\bibfield  {journal} {\bibinfo  {journal} {Nat.
	  mach. intell.}\ }\textbf {\bibinfo {volume} {3}},\ \bibinfo {pages} {316}
	  (\bibinfo {year} {2021})}\BibitemShut {NoStop}%
	\bibitem [{\citenamefont {Rossi}\ and\ \citenamefont {Ahmed}(2015)}]{nr}%
	  \BibitemOpen
	  \bibfield  {author} {\bibinfo {author} {\bibfnamefont {R.~A.}\ \bibnamefont
	  {Rossi}}\ and\ \bibinfo {author} {\bibfnamefont {N.~K.}\ \bibnamefont
	  {Ahmed}},\ }in\ \href {http://networkrepository.com} {\emph {\bibinfo
	  {booktitle} {Proceedings of the Twenty-Ninth AAAI Conference on Artificial
	  Intelligence}}}\ (\bibinfo {year} {2015})\BibitemShut {NoStop}%
\end{thebibliography}
\end{document}